\documentclass[twocolumn,aps,prd,nofootinbib]{revtex4-1}
\usepackage{graphicx,soul}
\usepackage{subfigure}
\usepackage{epsfig}
\usepackage{amstext,amsmath,amsfonts,amssymb,amsthm}
\usepackage{color}

\usepackage{verbatim}
\usepackage{lipsum}
\usepackage{color,array,wrapfig}
\usepackage{mathtools}
\usepackage{xcolor}
\usepackage[normalem]{ulem}
\def\be{\begin{equation}}
\def\ee{\end{equation}}
\def\bes{\begin{eqnarray}}
\def\ees{\end{eqnarray}}
\def\ba{\begin{align}}
\def\ea{\end{align}}
\def\bwt{\begin{widetext}}
\def\ewt{\end{widetext}}

\def\aa{\alpha}
\def\tt{\tau}

\def\aa{\aleph}
\def\tt{\overline{T}}

\newbox\one
\newbox\two
\long\def\loremlines#1{%
    \setbox\one=\vbox {%
       Test.\footnote{a footnote}%
      \lipsum\footnote{Another footnote.}%
     }
   \setbox\two=\vsplit\one to #1\baselineskip
   \unvbox\two}
\begin{document}
\title{Gravitational wave generation in a viable
scenario of inflationary magnetogenesis}

\title{Constraining models of Inflationary Magnetogenesis with NANOGrav}
\author{Ramkishor Sharma}
\email{ramkishor@iucaa.in}
\affiliation{IUCAA, Post Bag 4, Pune University Campus, Ganeshkhind, Pune$-$411007 India.}
\begin{abstract}
Generation of magnetic field during inflation can explain its presence over a wide range of scales in the Universe. In Ref.\cite{sharma2017}, we proposed a model to generate these fields during inflation. These fields have nonzero anisotropic stress which lead to the generation of a stochastic background of gravitational waves (GW) in the early universe. Here we show that for a scenario of magnetogenesis where reheating takes place around QCD epoch, this stochastic GW background lies in the $95\%$ confidence region of the GW signal probed by NANOGrav collaboration. This is the case when the generated electromagnetic field (EM) energy density is $3-10\%$ of the background energy density at the end of reheating. For this case, the values of magnetic field strength $B_0 \sim (3.8-6.9) \times 10^{-11}$G and its coherence length $\sim 30$ kpc at the present epoch. These values are for the models in which EM fields are of nonhelical nature. For the helical nature of the fields, these values are $B_0 \sim (1.1-1.9) \times 10^{-9}$G and its coherence length $\sim 0.8$ Mpc.
\end{abstract}
\maketitle
\section{Introduction} \label{introduction}
Magnetic fields have been observed over a wide range of scales in the universe from planets, stars to galactic and eaxtra-galactic scales \cite{r-beck,clarke,widrow,neronov,taylor2011}. These fields are assumed to be generated by the amplification of seed fields via flux freezing evolution followed by a turbulent dynamo mechanism \cite{kandu2019}. 
A number of scenarios of generation of 
seed magnetic fields have been suggested in literature such as generation during inflation \cite{turner-widrow, ratra, takahashi2005, shiv2005, martin-yokoyama, campanelli2008, rajeev2010, agullo2013,rajeev2013,caprini2014,kobayashi2014, atmjeet2014,sriram2015, Campanelli2015, 1475-7516-2015-03-040,arun2015,sriram2016,fujita2016,shinji2016,sumanta2018,fujita2019}, phase transitions \cite{vachaspati, Sigl:1996dm, kisslinger,qcd}, recombination, reionization 
and structure formation \cite{biermann,fenu,kandu1994,Zweibel2000,kulsrud1997}. 
The importance of inflationary scenarios of magnetic field generation as against other mechanisms lies in the fact that the former gives a natural way of generating fields coherent on large length scales. A popular model of inflationary generation involves coupling of a time dependent function with the usual 
electromagnetic (EM) action. In particular \citet{ratra} 
model takes the lagarangian density of the form $f^2 F^{\mu \nu} F_{\mu \nu}$ where $f$ is a function of inflaton field and $F_{\mu \nu}$ the electromagnetic field tensor. 
Although this model generates magnetic fields of sufficient strength to satisfy a number of observational constraints, 
it potentially suffers from back-reaction and strong coupling problems 
\cite{mukhanov2009}. It is also strongly constrained by the Schwinger effect which leads to the production of charged particles and arrests the growth of magnetic field \cite{kobayashi:2014}.

In a recent study \cite{sharma2017},
we have suggested a scenario in which these problems can be circumvented at the cost of having a low scale inflation. 
In this model, the coupling function $f$ increases during inflation starting from an initial value of 
unity and becomes very large at the end of inflation. Such an evolution of $f$ is free from the above mentioned problems. However, the coupling between the charges and EM field becomes very small at the end. To get back the standard EM theory we introduced a transition 
in the evolution of $f$ immediately after the end of inflation during which time it decreases back to unity at about reheating epoch and after that $f$ becomes a constant. During this post-inflationary era both electric and magnetic energy density 
increase. By demanding that EM energy density should remain below the background energy density, we obtained a bound on reheating and inflationary scales. Our models can generate both non-helical and helical magnetic fields and satisfy known observational constraints. They predict a blue spectrum for the magnetic field energy density peaked at small length scales, typically a fraction of
the Hubble radius at reheating
\cite{sharma2017,sharmahelical}. The generated field energy density 
can also be a significant fraction of the energy density of the
Universe at those epochs.

The anisotropic stresses associated with such primordial EM fields lead to the production of a stochastic gravitational wave background. In a recent study, we have estimated the produced GW spectrum in such a scenario of inflationary magnetogenesis. Recently, the North American Nanohertz Observatory for Gravitational Waves (NANOGrav) collaboration has reported evidence for a stochastic GW spectrum signal in the frequency range $[2.5 \times 10^{-9} Hz, 7.0 \times 10^{-8} Hz]$. Assuming that this signal is due to a stochastic background of GW, there have been various suggestions for their origin. These include mergers of super-massive black holes \cite{nanograv,nanogavastro1,nanogravastro2} or scenarios involving cosmic string \cite{nanogravstring1,nanogravstring2,nanogravstring3,bian2020}, primordial black holes \cite{nanopbh1,nanopbh2,nanopbh3,nanopbh4,soda2020,domenech2020}, phase transitions \cite{marek2020dec,addazi2020,nakai2020} and magneto-hydrodynamics turbulence during the QCD phase transition\cite{caprini2020} and others \cite{samanta2020,ratzinger2020,vagnozzi2020,pandey2020,ramberg2020,bhattacharya2020}. In this work, we focus on the GW spectrum produced in our model of inflationary magnetogenesis where reheating takes place around QCD epoch ($~150~\text{MeV}$) and compare the predicted signals with those reported by NANOGrav Collaboration.

The paper is organised as follows. In section \ref{gwenergyspectrum} we summarize the GW background generated in our models of inflationary magnetogenesis.
In section \ref{nano}, we compare the results of our model with the reported evidence of a stochastic background of GW by NANOGrav.
The last section \ref{conclusion} contains a discussion of our results and 
conclusions.

\section{Gravitational waves produced by electromagentic field generated during inflation}\label{gwenergyspectrum}
Gravitational waves are represented by the transverse trace-less part of the metric perturbations. Any source which has non zero transverse and trace-less part in its energy momentum tensor can lead to the production of gravitational waves in the early universe. In our case such a source is the electromagnetic field, generated during inflation and further during the reheating era. The stochastic GW spectrum results from this source was estimated in our earlier study \cite{sharmagw}. Here we give a summary relevant for the current work (see \citet{sharmagw} for details).

We consider a FLRW spacetime for the background geometry in the early universe. The metric including the tensor perturbations can be expressed as,
\begin{align*}
ds^2=a^2(\eta)(-d \eta^2 +(\delta_{ij}+2 h_{ij})dx^idx^j) .
\end{align*}
Here $x^i$ represents the comoving coordinates for the space dimensions and $\eta$ is the conformal time, $a(\eta)$ is the scale factor and $h_{ij}$ represents transverse trace-less part of the metric perturbations.
The spectrum of GW can be expressed in terms of the tensor perturbation as,
\begin{align}\label{gwenergydensity}
\frac{d\Omega_{GW}}{d \ln k}\Bigg|_0=\frac{k^3 a^2}{4(2 \pi)^3 G \rho_{c_0}}\sum_{\aa}\left(\Big|\frac{dh^{\aa}(k,\eta)}{d \eta}\Big|^2\right),
\end{align}
Here $d\Omega_{GW}(k)$ is the energy density in GW as a function of the closure density $\rho_c$, in a logarithmic interval ($d \ln k$) in wave number space. Also, $\aa$ represents the different polarisation state of GW and $\aa={T,\times}$ or $\aa={+,-}$ depending on whether it is linearly or circularly polarised.
The evolution of the tensor perturbation ($h_{ij}$) in presence of a source, is given by linearised Einstein's equation,
\begin{align}\label{gweom}
h''_{\aa}+\frac{2 a'}{a} h'_{\aa}+ k^2 h_{\aa} =8 \pi G a^2 (\rho+p) \Pi_{\aa} .
\end{align}
where $\Pi_{\aa}$ is defined as $\Pi_{\aa}\equiv[1/(\rho+p)] \tt_{\aa}^{TT}$ and $\tt_{\aa}^{TT}$ is the transverse trace-less part of the energy-momentum tensor of the source. 

In our earlier studies Ref.\cite{sharma2017, sharmahelical}, to address the strong coupling and back-reaction problems in  $f^2 F^{\mu \nu} F_{\mu \nu}$ type models of inflationary inflationary magnetogenesis, we have taken a particular evolution of the coupling function, $f$ which evolves with time both during as well as post inflation till reheating. This function increases during inflation and transits to a decaying phase post inflation. We have assumed that Universe evolves as in a matter dominated era between the end of inflation and the beginning of reheating. After this matter dominated era, reheating takes place and standard radiation dominance starts. During inflation the magnetic field spectrum is scale invariant but the strength is very low compared to the background energy density because of the low energy scale of inflation. In the post inflationary era when coupling function, $f$ decreases, the scale invariant contribution to the magnetic spectrum decreases but contribution from the next order gets amplified on the superhorizon scales. This post inflationary era ends when the EM energy density is $\epsilon$ times the background energy density and after this reheating takes place and EM energy density evolves like radiation. The magnetic field spectrum generated in our model 
is a blue spectrum, $d \tilde{\rho}_{B}(k,\eta)/d \ln k
\propto k^4$, where $\tilde{\rho}_B$ is the comoving magnetic
energy density. The EM fields generated can lead to the production of GW. The main contribution to the GW energy spectrum is during the end phase of the post inflationary matter dominated era. During this era both electric and magnetic fields contribute to the production of GW. However after reheating, electric fields get shorted out because of the large conductivity of the universe and only magnetic field contributes to the production of GW. 

In our case, the GW spectrum depends upon the expectation value of $\tt_{ij}(\eta_1) \tt^{ij}(\eta_2)$ at unequal times and Green function which takes the form $\cos(k(\eta_1-\eta_2)/(\eta_1 \eta_2))$ in
the sub horizon limit in a radiation dominated universe. Further, we write the former as the expectation value of $\tt_{ij} \tt^{ij}$ at equal time and an unequal time correlation function of the EM fields. The GW spectrum has been obtained by numerically solving the expressions and the details are provided in section (IV) in \citet{sharmagw} and it can be summarised as follows. The generated GW spectrum rises with wavenumber $k$ as $d \Omega_{GW}/d \ln(k) \propto k^3$,
at low wavenumbers. It remains almost $k^3$ until the wavenumber $k=k_{peak}$ for $\epsilon=1$, where $\epsilon$ denotes the fraction of EM energy density to the background energy density at reheating. However, for $\epsilon=10^{-2}$, it changes to a shallower spectrum compared to $k^3$. The GW spectrum then falls for the modes $k>k_{peak}$ as $d \Omega_{GW}/d \ln(k) \propto k^{-5/3}$ for $\epsilon=1$ and $d \Omega_{GW}/d \ln(k) \propto k^{-8/3}$ for $\epsilon=10^{-2}$. This change in the slope of the spectrum for different $\epsilon$ could arise due to the turbulence correlation time being longer for a smaller $\epsilon$.

\section{Comparison with the NANOGrav Signal}\label{nano}
The NANOGrav collaboration has recently reported evidence for a stochastic background of GW using 12.5 years of timing residual data set of pulsars \cite{nanograv}. The time residual cross power spectral density versus frequency data set has 30 frequency components in the range $[2.5 \times 10^{-9} Hz, 7.0 \times 10^{-8} Hz]$. In the NANOGrav collaboration paper\cite{nanograv}, the time residual cross power spectral density has been modelled as a simple power law and a broken power law in frequency and the $68\%$ and $95\%$ confidence regimes for the amplitude and the spectral index using the lowest five frequency components have been determined. For our analysis, it is the broken power law case that is relevant. For this case, we convert the modelled time residual cross power spectral density in terms of GW density fraction ($\Omega_{GW}$) using Eq.(29) in Ref.\cite{moore2014},
\begin{equation}
    \Omega_{GW}=\frac{2 \pi^2 A_{GWB}^2 f_{yr}^{2}}{3 H_0^2} \left(\frac{f}{f_{yr}}\right)^{5-\gamma}\left(1+\left(\frac{f}{f_{bend}}\right)^{\frac{1}{\kappa}}\right)^{\kappa(\gamma-5-\delta)} .
\end{equation}
Here $A_{GWB}$ is the characteristic strain at $f=f_{yr}$, $H_0$ is the value of Hubble parameter today, $\gamma$ and $\delta$ are the power law index at frequencies lower and higher than $f_{bend}=1.2\times10^{-8}$Hz, respectively, and $\kappa$ controls the smoothness of the transition. In Ref.\cite{nanograv} $\delta$ is taken to be zero and $\kappa=0.1$. The $95\%$ confidence contour in terms of the frequency spectrum for density fraction and frequency is shown in Fig.(1) in pink colour. To determine this contour, we plot all the value of $A_{GWB}$ and $\gamma$ for the $95\%$ confidence region and extract the maximum covered area. For $95\%$ confidence region, $\gamma\in(3.1,6.7)$. As it is evident from the left panel of Fig.(1) in Ref.\cite{neronov} that the time residual has very large spread for $f>f_{bend}$ compared to the value for $f<f_{bend}$. Therefore, we do not include the confidence contour for $f>f_{bend}$ while comparing with the GW signal obtained in a model. We plot the resulting GW spectrum for our model in the same figure for different values of the ratio of EM energy density to the background energy density ($\epsilon$), and a reheating temperature $T_R=150$ MeV.

From Fig.(1), we conclude that the GW produced in our model, for a scenario in which reheating temperature is $T_R\sim150$ MeV, lies within the $95\%$ confidence regime of parameter space of the signal, modelled as a broken power law, for $\epsilon=(0.03,0.1)$. 

\begin{widetext}
\begin{center}
\begin{figure}\label{fig1}
\includegraphics[scale=0.5]{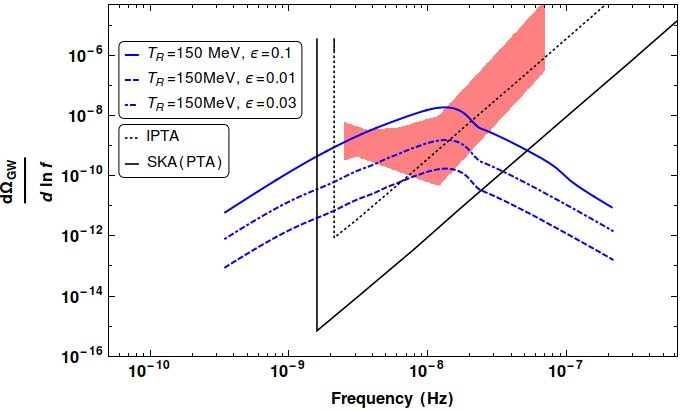}
\caption{In this figure, density fraction of gravitational waves in logarithmic frequency interval with frequency is shown. The blue curves are the GW spectrum obtained in our model of inflationary magnetogenesis for a scenario where reheating temperature $T_R$=150 MeV for different value of the ratio of EM and background energy density ($\epsilon$). Pink colour region is $95\%$ confidence region of the parameter space of the GW signal modelled as broken power law by NANOGrav collaboration \cite{nanograv}. Solid and dotted black curves represent the sensitivity curve of the international pulsar timing array and pulsar timing array with the upcoming mission square kilometer array (SKA), respectively \citet{moore2014}.}
\end{figure}
\end{center}
\end{widetext}

Further, we estimate the magnetic field strength today consistent with the signal seen by the NANOGrav collaboration in \cite{nanograv}. After the generation, apart from the dilution due to the adiabatic expansion, magnetic fields also undergo turbulent decay due to the non-linear processing. Including these effects, the magnetic field and its coherence length at the present epoch can be related to its strength  and coherence length at the epoch of generation as follows,
\begin{equation}
    B_0\approx\sqrt{\frac{\epsilon}{2}} \left(\frac{a_0}{a_g}\right)^{-2}\left(\frac{a_m}{a_g}\right)^{-p}, L_{c0}=L_c\left(\frac{a_0}{a_g}\right) \left(\frac{a_m}{a_g}\right)^q
\end{equation}
Here, $p=(n+3)/(n+5)$ and $q=2/(n+5)$, $n$ is defined such that power spectrum of the magnetic field, $P_{B}\propto k^n$. $a_0$, $a_m$ and $a_g$ represent the value of the scale factor at the present epoch, matter-radiation equality, and end of reheating, respectively and $L_c$ is the coherence length of the magnetic field at the end reheating. Using the above expression, we get $B_0 \sim (3.8-6.9) \times 10^{-11}$G for and its coherence length $\sim 30$ kpc for the case when the GW spectrum from EM fields anisotropic stresses is consistent with the signal found by NANOGrav collaboration. To calculate these numbers, we take $p=0.5~\text{and}~q=0.5$ suggested by numerical simulation for the evolution of magnetic field in the early universe \cite{axel,axel2017,zrake}.

The strength of the generated GW spectrum in the case of helical EM fields is similar to the nonhelical case \cite{sharmagw}.
However, in the case of helical EM fields, the generated GW spectrum is circularly polarised while it is unpolarised
when the EM fields are nonhelical. Assuming that fully helical EM fields explain the NANOGrav signal,  we get $B_0 \sim (1.1-1.9) \times 10^{-9}$G and its coherence length $\sim (0.8)$ Mpc. Here we use $p=-1/3~\text{and}~q=2/3$ for the evolution of helical magnetic field in the early universe \cite{jedamzik,tina2013,axel2016}. Observational limits on CMB non-Gaussianity from the Planck mission set an upper limit of $B_0\le0.6$nG on the present value of the primordial cosmic magnetic field \cite{trivedi2013} and this limit has been obtained for a scale invariant spectrum of magnetic field. 


\section{Discussion and Conclusion}\label{conclusion}
The origin of magnetic fields observed in large scale structures is a question of astrophysical interest. Generation of these fields during inflation is an interesting possibility. These fields can be generated in a scenario suggested in \citet{sharma2017} which is free from the difficulties raised in the literature. The generated magnetic fields have non zero anisotopic stresses which lead to the production of stochastic background of gravitational waves. If reheating in our scenario takes place around the QCD epoch, the resulting GW background can be interpreted as leading to the signal inferred in NANOGrav 12.5 year data. This requires the EM field energy density to be in the range of $3\%$ to $10\%$ of the background energy density at generation. The magnetic fields consistent with the NANOGrav signal has a present day strength $B_0 \sim (3.8-6.9) \times 10^{-11}$G and their coherence length $\sim 30$ kpc when the EM fields are of non-helical nature. These values change to  $B_0 \sim (31.1-1.9) \times 10^{-9}$G and $\sim 0.8$ Mpc for the helical case.

In this work, we only consider the scenario where reheating temperature, $T_R=150$ MeV. However, scenarios where reheating temperature ranges above 10's of MeV (as required to obtain standard big bang nucleosynthesis \cite{bbn}) to GeV range can also be constrained by the pulsar timing arrays. For these reheating temperature scales, the nature of the GW spectrum remains the same, except that the signal shown in Fig.(1) shifts towards left for $T_R<150$ MeV and towards the right side for $T_R>150$MeV. A future data set for timing residuals of pulsars with observation for more years, may help in better constraining such models of inflationary magnetogenesis. If the data set turns out to favor a broken power law and follows the power-law slopes consistent with our model's prediction, then it uniquely fixes the scale of reheating in our model with the help of the peak frequency of the spectrum. The nature of resulting GW signal in our model is non-Gaussian; this property may help in distinguish these models from other models of GW generation in the early Universe \cite{bartolo2018}.


\section*{Acknowledgments}
The author would like to thank Prof. Kandaswamy Subramanian and Prof. T. R. Seshadri for several useful discussions and for providing valuable suggestions on the manuscript. The author is thankful to Prof. Bhal Chandra Joshi for the discussion and helpful insights on NANOGrav results.
\bibliographystyle{apsrev4-1}
\bibliography{references}

\begin{thebibliography}{74}%
\makeatletter
\providecommand \@ifxundefined [1]{%
 \@ifx{#1\undefined}
}%
\providecommand \@ifnum [1]{%
 \ifnum #1\expandafter \@firstoftwo
 \else \expandafter \@secondoftwo
 \fi
}%
\providecommand \@ifx [1]{%
 \ifx #1\expandafter \@firstoftwo
 \else \expandafter \@secondoftwo
 \fi
}%
\providecommand \natexlab [1]{#1}%
\providecommand \enquote  [1]{``#1''}%
\providecommand \bibnamefont  [1]{#1}%
\providecommand \bibfnamefont [1]{#1}%
\providecommand \citenamefont [1]{#1}%
\providecommand \href@noop [0]{\@secondoftwo}%
\providecommand \href [0]{\begingroup \@sanitize@url \@href}%
\providecommand \@href[1]{\@@startlink{#1}\@@href}%
\providecommand \@@href[1]{\endgroup#1\@@endlink}%
\providecommand \@sanitize@url [0]{\catcode `\\12\catcode `\$12\catcode
  `\&12\catcode `\#12\catcode `\^12\catcode `\_12\catcode `\%12\relax}%
\providecommand \@@startlink[1]{}%
\providecommand \@@endlink[0]{}%
\providecommand \url  [0]{\begingroup\@sanitize@url \@url }%
\providecommand \@url [1]{\endgroup\@href {#1}{\urlprefix }}%
\providecommand \urlprefix  [0]{URL }%
\providecommand \Eprint [0]{\href }%
\providecommand \doibase [0]{http://dx.doi.org/}%
\providecommand \selectlanguage [0]{\@gobble}%
\providecommand \bibinfo  [0]{\@secondoftwo}%
\providecommand \bibfield  [0]{\@secondoftwo}%
\providecommand \translation [1]{[#1]}%
\providecommand \BibitemOpen [0]{}%
\providecommand \bibitemStop [0]{}%
\providecommand \bibitemNoStop [0]{.\EOS\space}%
\providecommand \EOS [0]{\spacefactor3000\relax}%
\providecommand \BibitemShut  [1]{\csname bibitem#1\endcsname}%
\let\auto@bib@innerbib\@empty
\bibitem [{\citenamefont {{Sharma}}\ \emph {et~al.}(2017)\citenamefont
  {{Sharma}}, \citenamefont {{Jagannathan}}, \citenamefont {{Seshadri}},\ and\
  \citenamefont {{Subramanian}}}]{sharma2017}%
  \BibitemOpen
  \bibfield  {author} {\bibinfo {author} {\bibfnamefont {R.}~\bibnamefont
  {{Sharma}}}, \bibinfo {author} {\bibfnamefont {S.}~\bibnamefont
  {{Jagannathan}}}, \bibinfo {author} {\bibfnamefont {T.~R.}\ \bibnamefont
  {{Seshadri}}}, \ and\ \bibinfo {author} {\bibfnamefont {K.}~\bibnamefont
  {{Subramanian}}},\ }\href {\doibase 10.1103/PhysRevD.96.083511} {\bibfield
  {journal} {\bibinfo  {journal} {\prd}\ }\textbf {\bibinfo {volume} {96}},\
  \bibinfo {eid} {083511} (\bibinfo {year} {2017})},\ \Eprint
  {http://arxiv.org/abs/1708.08119} {arXiv:1708.08119} \BibitemShut {NoStop}%
\bibitem [{\citenamefont {Beck}(2001)}]{r-beck}%
  \BibitemOpen
  \bibfield  {author} {\bibinfo {author} {\bibfnamefont {R.}~\bibnamefont
  {Beck}},\ }\href {\doibase 10.1023/A:1013805401252} {\bibfield  {journal}
  {\bibinfo  {journal} {Space Sci. Rev.}\ }\textbf {\bibinfo {volume} {99}},\
  \bibinfo {pages} {243} (\bibinfo {year} {2001})},\ \Eprint
  {http://arxiv.org/abs/astro-ph/0012402} {arXiv:astro-ph/0012402 [astro-ph]}
  \BibitemShut {NoStop}%
\bibitem [{\citenamefont {Clarke}\ \emph {et~al.}(2001)\citenamefont {Clarke},
  \citenamefont {Kronberg},\ and\ \citenamefont {Boehringer}}]{clarke}%
  \BibitemOpen
  \bibfield  {author} {\bibinfo {author} {\bibfnamefont {T.~E.}\ \bibnamefont
  {Clarke}}, \bibinfo {author} {\bibfnamefont {P.~P.}\ \bibnamefont
  {Kronberg}}, \ and\ \bibinfo {author} {\bibfnamefont {H.}~\bibnamefont
  {Boehringer}},\ }\href {\doibase 10.1086/318896} {\bibfield  {journal}
  {\bibinfo  {journal} {Astrophys. J.}\ }\textbf {\bibinfo {volume} {547}},\
  \bibinfo {pages} {L111} (\bibinfo {year} {2001})},\ \Eprint
  {http://arxiv.org/abs/astro-ph/0011281} {arXiv:astro-ph/0011281 [astro-ph]}
  \BibitemShut {NoStop}%
\bibitem [{\citenamefont {{Widrow}}(2002)}]{widrow}%
  \BibitemOpen
  \bibfield  {author} {\bibinfo {author} {\bibfnamefont {L.~M.}\ \bibnamefont
  {{Widrow}}},\ }\href {\doibase 10.1103/RevModPhys.74.775} {\bibfield
  {journal} {\bibinfo  {journal} {Reviews of Modern Physics}\ }\textbf
  {\bibinfo {volume} {74}},\ \bibinfo {pages} {775} (\bibinfo {year} {2002})},\
  \Eprint {http://arxiv.org/abs/astro-ph/0207240} {astro-ph/0207240}
  \BibitemShut {NoStop}%
\bibitem [{\citenamefont {{Neronov}}\ and\ \citenamefont
  {{Vovk}}(2010)}]{neronov}%
  \BibitemOpen
  \bibfield  {author} {\bibinfo {author} {\bibfnamefont {A.}~\bibnamefont
  {{Neronov}}}\ and\ \bibinfo {author} {\bibfnamefont {I.}~\bibnamefont
  {{Vovk}}},\ }\href {\doibase 10.1126/science.1184192} {\bibfield  {journal}
  {\bibinfo  {journal} {Science}\ }\textbf {\bibinfo {volume} {328}},\ \bibinfo
  {pages} {73} (\bibinfo {year} {2010})},\ \Eprint
  {http://arxiv.org/abs/1006.3504} {arXiv:1006.3504 [astro-ph.HE]} \BibitemShut
  {NoStop}%
\bibitem [{\citenamefont {Taylor}\ \emph {et~al.}(2011)\citenamefont {Taylor},
  \citenamefont {Vovk},\ and\ \citenamefont {Neronov}}]{taylor2011}%
  \BibitemOpen
  \bibfield  {author} {\bibinfo {author} {\bibfnamefont {A.~M.}\ \bibnamefont
  {Taylor}}, \bibinfo {author} {\bibfnamefont {I.}~\bibnamefont {Vovk}}, \ and\
  \bibinfo {author} {\bibfnamefont {A.}~\bibnamefont {Neronov}},\ }\href
  {\doibase 10.1051/0004-6361/201116441} {\bibfield  {journal} {\bibinfo
  {journal} {Astron. Astrophys.}\ }\textbf {\bibinfo {volume} {529}},\ \bibinfo
  {pages} {A144} (\bibinfo {year} {2011})},\ \Eprint
  {http://arxiv.org/abs/1101.0932} {arXiv:1101.0932 [astro-ph.HE]} \BibitemShut
  {NoStop}%
\bibitem [{\citenamefont {Subramanian}(2019)}]{kandu2019}%
  \BibitemOpen
  \bibfield  {author} {\bibinfo {author} {\bibfnamefont {K.}~\bibnamefont
  {Subramanian}},\ }\href {\doibase 10.3390/galaxies7020047} {\bibfield
  {journal} {\bibinfo  {journal} {Galaxies}\ }\textbf {\bibinfo {volume} {7}},\
  \bibinfo {pages} {47} (\bibinfo {year} {2019})},\ \Eprint
  {http://arxiv.org/abs/1903.03744} {arXiv:1903.03744 [astro-ph.CO]}
  \BibitemShut {NoStop}%
\bibitem [{\citenamefont {{Turner}}\ and\ \citenamefont
  {{Widrow}}(1988)}]{turner-widrow}%
  \BibitemOpen
  \bibfield  {author} {\bibinfo {author} {\bibfnamefont {M.~S.}\ \bibnamefont
  {{Turner}}}\ and\ \bibinfo {author} {\bibfnamefont {L.~M.}\ \bibnamefont
  {{Widrow}}},\ }\href {\doibase 10.1103/PhysRevD.37.2743} {\bibfield
  {journal} {\bibinfo  {journal} {\prd}\ }\textbf {\bibinfo {volume} {37}},\
  \bibinfo {pages} {2743} (\bibinfo {year} {1988})}\BibitemShut {NoStop}%
\bibitem [{\citenamefont {Ratra}(1992)}]{ratra}%
  \BibitemOpen
  \bibfield  {author} {\bibinfo {author} {\bibfnamefont {B.}~\bibnamefont
  {Ratra}},\ }\href {\doibase 10.1086/186384} {\bibfield  {journal} {\bibinfo
  {journal} {Astrophys. J.}\ }\textbf {\bibinfo {volume} {391}},\ \bibinfo
  {pages} {L1} (\bibinfo {year} {1992})}\BibitemShut {NoStop}%
\bibitem [{\citenamefont {{Takahashi}}\ \emph {et~al.}(2005)\citenamefont
  {{Takahashi}}, \citenamefont {{Ichiki}}, \citenamefont {{Ohno}},\ and\
  \citenamefont {{Hanayama}}}]{takahashi2005}%
  \BibitemOpen
  \bibfield  {author} {\bibinfo {author} {\bibfnamefont {K.}~\bibnamefont
  {{Takahashi}}}, \bibinfo {author} {\bibfnamefont {K.}~\bibnamefont
  {{Ichiki}}}, \bibinfo {author} {\bibfnamefont {H.}~\bibnamefont {{Ohno}}}, \
  and\ \bibinfo {author} {\bibfnamefont {H.}~\bibnamefont {{Hanayama}}},\
  }\href {\doibase 10.1103/PhysRevLett.95.121301} {\bibfield  {journal}
  {\bibinfo  {journal} {Physical Review Letters}\ }\textbf {\bibinfo {volume}
  {95}},\ \bibinfo {eid} {121301} (\bibinfo {year} {2005})},\ \Eprint
  {http://arxiv.org/abs/astro-ph/0502283} {astro-ph/0502283} \BibitemShut
  {NoStop}%
\bibitem [{\citenamefont {Gopal}\ and\ \citenamefont {Sethi}(2005)}]{shiv2005}%
  \BibitemOpen
  \bibfield  {author} {\bibinfo {author} {\bibfnamefont {R.}~\bibnamefont
  {Gopal}}\ and\ \bibinfo {author} {\bibfnamefont {S.}~\bibnamefont {Sethi}},\
  }\href {\doibase 10.1111/j.1365-2966.2005.09442.x} {\bibfield  {journal}
  {\bibinfo  {journal} {Mon. Not. Roy. Astron. Soc.}\ }\textbf {\bibinfo
  {volume} {363}},\ \bibinfo {pages} {521} (\bibinfo {year} {2005})},\ \Eprint
  {http://arxiv.org/abs/astro-ph/0411170} {arXiv:astro-ph/0411170 [astro-ph]}
  \BibitemShut {NoStop}%
\bibitem [{\citenamefont {{Martin}}\ and\ \citenamefont
  {{Yokoyama}}(2008)}]{martin-yokoyama}%
  \BibitemOpen
  \bibfield  {author} {\bibinfo {author} {\bibfnamefont {J.}~\bibnamefont
  {{Martin}}}\ and\ \bibinfo {author} {\bibfnamefont {J.}~\bibnamefont
  {{Yokoyama}}},\ }\href {\doibase 10.1088/1475-7516/2008/01/025} {\bibfield
  {journal} {\bibinfo  {journal} {Journal of Cosmology and Astroparticle
  Physics}\ }\textbf {\bibinfo {volume} {1}},\ \bibinfo {eid} {025} (\bibinfo
  {year} {2008})},\ \Eprint {http://arxiv.org/abs/0711.4307} {arXiv:0711.4307}
  \BibitemShut {NoStop}%
\bibitem [{\citenamefont {{Campanelli}}\ \emph {et~al.}(2008)\citenamefont
  {{Campanelli}}, \citenamefont {{Cea}}, \citenamefont {{Fogli}},\ and\
  \citenamefont {{Tedesco}}}]{campanelli2008}%
  \BibitemOpen
  \bibfield  {author} {\bibinfo {author} {\bibfnamefont {L.}~\bibnamefont
  {{Campanelli}}}, \bibinfo {author} {\bibfnamefont {P.}~\bibnamefont {{Cea}}},
  \bibinfo {author} {\bibfnamefont {G.~L.}\ \bibnamefont {{Fogli}}}, \ and\
  \bibinfo {author} {\bibfnamefont {L.}~\bibnamefont {{Tedesco}}},\ }\href
  {\doibase 10.1103/PhysRevD.77.043001} {\bibfield  {journal} {\bibinfo
  {journal} {\prd}\ }\textbf {\bibinfo {volume} {77}},\ \bibinfo {eid} {043001}
  (\bibinfo {year} {2008})},\ \Eprint {http://arxiv.org/abs/0710.2993}
  {arXiv:0710.2993} \BibitemShut {NoStop}%
\bibitem [{\citenamefont {Durrer}\ \emph {et~al.}(2011)\citenamefont {Durrer},
  \citenamefont {Hollenstein},\ and\ \citenamefont {Jain}}]{rajeev2010}%
  \BibitemOpen
  \bibfield  {author} {\bibinfo {author} {\bibfnamefont {R.}~\bibnamefont
  {Durrer}}, \bibinfo {author} {\bibfnamefont {L.}~\bibnamefont {Hollenstein}},
  \ and\ \bibinfo {author} {\bibfnamefont {R.~K.}\ \bibnamefont {Jain}},\
  }\href {\doibase 10.1088/1475-7516/2011/03/037} {\bibfield  {journal}
  {\bibinfo  {journal} {JCAP}\ }\textbf {\bibinfo {volume} {1103}},\ \bibinfo
  {pages} {037} (\bibinfo {year} {2011})},\ \Eprint
  {http://arxiv.org/abs/1005.5322} {arXiv:1005.5322 [astro-ph.CO]} \BibitemShut
  {NoStop}%
\bibitem [{\citenamefont {Agullo}\ and\ \citenamefont
  {Navarro-Salas}(2013)}]{agullo2013}%
  \BibitemOpen
  \bibfield  {author} {\bibinfo {author} {\bibfnamefont {I.}~\bibnamefont
  {Agullo}}\ and\ \bibinfo {author} {\bibfnamefont {J.}~\bibnamefont
  {Navarro-Salas}},\ }\href@noop {} {\  (\bibinfo {year} {2013})},\ \Eprint
  {http://arxiv.org/abs/1309.3435} {arXiv:1309.3435 [gr-qc]} \BibitemShut
  {NoStop}%
\bibitem [{\citenamefont {{Ferreira}}\ \emph {et~al.}(2013)\citenamefont
  {{Ferreira}}, \citenamefont {{Jain}},\ and\ \citenamefont
  {{Sloth}}}]{rajeev2013}%
  \BibitemOpen
  \bibfield  {author} {\bibinfo {author} {\bibfnamefont {R.~J.~Z.}\
  \bibnamefont {{Ferreira}}}, \bibinfo {author} {\bibfnamefont {R.~K.}\
  \bibnamefont {{Jain}}}, \ and\ \bibinfo {author} {\bibfnamefont {M.~S.}\
  \bibnamefont {{Sloth}}},\ }\href {\doibase 10.1088/1475-7516/2013/10/004}
  {\bibfield  {journal} {\bibinfo  {journal} {Journal of Cosmology and
  Astroparticle Physics}\ }\textbf {\bibinfo {volume} {10}},\ \bibinfo {eid}
  {004} (\bibinfo {year} {2013})},\ \Eprint {http://arxiv.org/abs/1305.7151}
  {arXiv:1305.7151 [astro-ph.CO]} \BibitemShut {NoStop}%
\bibitem [{\citenamefont {Caprini}\ and\ \citenamefont
  {Sorbo}(2014)}]{caprini2014}%
  \BibitemOpen
  \bibfield  {author} {\bibinfo {author} {\bibfnamefont {C.}~\bibnamefont
  {Caprini}}\ and\ \bibinfo {author} {\bibfnamefont {L.}~\bibnamefont
  {Sorbo}},\ }\href {\doibase 10.1088/1475-7516/2014/10/056} {\bibfield
  {journal} {\bibinfo  {journal} {JCAP}\ }\textbf {\bibinfo {volume} {1410}},\
  \bibinfo {pages} {056} (\bibinfo {year} {2014})},\ \Eprint
  {http://arxiv.org/abs/1407.2809} {arXiv:1407.2809 [astro-ph.CO]} \BibitemShut
  {NoStop}%
\bibitem [{\citenamefont {Kobayashi}(2014)}]{kobayashi2014}%
  \BibitemOpen
  \bibfield  {author} {\bibinfo {author} {\bibfnamefont {T.}~\bibnamefont
  {Kobayashi}},\ }\href {http://stacks.iop.org/1475-7516/2014/i=05/a=040}
  {\bibfield  {journal} {\bibinfo  {journal} {Journal of Cosmology and
  Astroparticle Physics}\ }\textbf {\bibinfo {volume} {2014}},\ \bibinfo
  {pages} {040} (\bibinfo {year} {2014})}\BibitemShut {NoStop}%
\bibitem [{\citenamefont {{Atmjeet}}\ \emph {et~al.}(2014)\citenamefont
  {{Atmjeet}}, \citenamefont {{Pahwa}}, \citenamefont {{Seshadri}},\ and\
  \citenamefont {{Subramanian}}}]{atmjeet2014}%
  \BibitemOpen
  \bibfield  {author} {\bibinfo {author} {\bibfnamefont {K.}~\bibnamefont
  {{Atmjeet}}}, \bibinfo {author} {\bibfnamefont {I.}~\bibnamefont {{Pahwa}}},
  \bibinfo {author} {\bibfnamefont {T.~R.}\ \bibnamefont {{Seshadri}}}, \ and\
  \bibinfo {author} {\bibfnamefont {K.}~\bibnamefont {{Subramanian}}},\ }\href
  {\doibase 10.1103/PhysRevD.89.063002} {\bibfield  {journal} {\bibinfo
  {journal} {\prd}\ }\textbf {\bibinfo {volume} {89}},\ \bibinfo {eid} {063002}
  (\bibinfo {year} {2014})},\ \Eprint {http://arxiv.org/abs/1312.5815}
  {arXiv:1312.5815 [astro-ph.CO]} \BibitemShut {NoStop}%
\bibitem [{\citenamefont {Sriramkumar}\ \emph {et~al.}(2015)\citenamefont
  {Sriramkumar}, \citenamefont {Atmjeet},\ and\ \citenamefont
  {Jain}}]{sriram2015}%
  \BibitemOpen
  \bibfield  {author} {\bibinfo {author} {\bibfnamefont {L.}~\bibnamefont
  {Sriramkumar}}, \bibinfo {author} {\bibfnamefont {K.}~\bibnamefont
  {Atmjeet}}, \ and\ \bibinfo {author} {\bibfnamefont {R.~K.}\ \bibnamefont
  {Jain}},\ }\href {\doibase 10.1088/1475-7516/2015/09/010} {\bibfield
  {journal} {\bibinfo  {journal} {JCAP}\ }\textbf {\bibinfo {volume} {1509}},\
  \bibinfo {pages} {010} (\bibinfo {year} {2015})},\ \Eprint
  {http://arxiv.org/abs/1504.06853} {arXiv:1504.06853 [astro-ph.CO]}
  \BibitemShut {NoStop}%
\bibitem [{\citenamefont {Campanelli}(2015)}]{Campanelli2015}%
  \BibitemOpen
  \bibfield  {author} {\bibinfo {author} {\bibfnamefont {L.}~\bibnamefont
  {Campanelli}},\ }\href {\doibase 10.1140/epjc/s10052-015-3510-x} {\bibfield
  {journal} {\bibinfo  {journal} {The European Physical Journal C}\ }\textbf
  {\bibinfo {volume} {75}},\ \bibinfo {pages} {278} (\bibinfo {year}
  {2015})}\BibitemShut {NoStop}%
\bibitem [{\citenamefont {Tasinato}(2015)}]{1475-7516-2015-03-040}%
  \BibitemOpen
  \bibfield  {author} {\bibinfo {author} {\bibfnamefont {G.}~\bibnamefont
  {Tasinato}},\ }\href {http://stacks.iop.org/1475-7516/2015/i=03/a=040}
  {\bibfield  {journal} {\bibinfo  {journal} {Journal of Cosmology and
  Astroparticle Physics}\ }\textbf {\bibinfo {volume} {2015}},\ \bibinfo
  {pages} {040} (\bibinfo {year} {2015})}\BibitemShut {NoStop}%
\bibitem [{\citenamefont {Bhatt}\ and\ \citenamefont
  {Pandey}(2016)}]{arun2015}%
  \BibitemOpen
  \bibfield  {author} {\bibinfo {author} {\bibfnamefont {J.~R.}\ \bibnamefont
  {Bhatt}}\ and\ \bibinfo {author} {\bibfnamefont {A.~K.}\ \bibnamefont
  {Pandey}},\ }\href {\doibase 10.1103/PhysRevD.94.043536} {\bibfield
  {journal} {\bibinfo  {journal} {Phys. Rev.}\ }\textbf {\bibinfo {volume}
  {D94}},\ \bibinfo {pages} {043536} (\bibinfo {year} {2016})},\ \Eprint
  {http://arxiv.org/abs/1503.01878} {arXiv:1503.01878 [astro-ph.CO]}
  \BibitemShut {NoStop}%
\bibitem [{\citenamefont {Chowdhury}\ \emph {et~al.}(2016)\citenamefont
  {Chowdhury}, \citenamefont {Sriramkumar},\ and\ \citenamefont
  {Jain}}]{sriram2016}%
  \BibitemOpen
  \bibfield  {author} {\bibinfo {author} {\bibfnamefont {D.}~\bibnamefont
  {Chowdhury}}, \bibinfo {author} {\bibfnamefont {L.}~\bibnamefont
  {Sriramkumar}}, \ and\ \bibinfo {author} {\bibfnamefont {R.~K.}\ \bibnamefont
  {Jain}},\ }\href {\doibase 10.1103/PhysRevD.94.083512} {\bibfield  {journal}
  {\bibinfo  {journal} {Phys. Rev.}\ }\textbf {\bibinfo {volume} {D94}},\
  \bibinfo {pages} {083512} (\bibinfo {year} {2016})},\ \Eprint
  {http://arxiv.org/abs/1604.02143} {arXiv:1604.02143 [gr-qc]} \BibitemShut
  {NoStop}%
\bibitem [{\citenamefont {Fujita}\ and\ \citenamefont
  {Namba}(2016)}]{fujita2016}%
  \BibitemOpen
  \bibfield  {author} {\bibinfo {author} {\bibfnamefont {T.}~\bibnamefont
  {Fujita}}\ and\ \bibinfo {author} {\bibfnamefont {R.}~\bibnamefont {Namba}},\
  }\href {\doibase 10.1103/PhysRevD.94.043523} {\bibfield  {journal} {\bibinfo
  {journal} {Phys. Rev.}\ }\textbf {\bibinfo {volume} {D94}},\ \bibinfo {pages}
  {043523} (\bibinfo {year} {2016})},\ \Eprint
  {http://arxiv.org/abs/1602.05673} {arXiv:1602.05673 [astro-ph.CO]}
  \BibitemShut {NoStop}%
\bibitem [{\citenamefont {Mukohyama}(2016)}]{shinji2016}%
  \BibitemOpen
  \bibfield  {author} {\bibinfo {author} {\bibfnamefont {S.}~\bibnamefont
  {Mukohyama}},\ }\href {\doibase 10.1103/PhysRevD.94.121302} {\bibfield
  {journal} {\bibinfo  {journal} {Phys. Rev.}\ }\textbf {\bibinfo {volume}
  {D94}},\ \bibinfo {pages} {121302} (\bibinfo {year} {2016})},\ \Eprint
  {http://arxiv.org/abs/1607.07041} {arXiv:1607.07041 [hep-th]} \BibitemShut
  {NoStop}%
\bibitem [{\citenamefont {Chakraborty}\ \emph {et~al.}(2018)\citenamefont
  {Chakraborty}, \citenamefont {Pal},\ and\ \citenamefont
  {SenGupta}}]{sumanta2018}%
  \BibitemOpen
  \bibfield  {author} {\bibinfo {author} {\bibfnamefont {S.}~\bibnamefont
  {Chakraborty}}, \bibinfo {author} {\bibfnamefont {S.}~\bibnamefont {Pal}}, \
  and\ \bibinfo {author} {\bibfnamefont {S.}~\bibnamefont {SenGupta}},\
  }\href@noop {} {\  (\bibinfo {year} {2018})},\ \Eprint
  {http://arxiv.org/abs/1810.03478} {arXiv:1810.03478 [gr-qc]} \BibitemShut
  {NoStop}%
\bibitem [{\citenamefont {Fujita}\ and\ \citenamefont
  {Durrer}(2019)}]{fujita2019}%
  \BibitemOpen
  \bibfield  {author} {\bibinfo {author} {\bibfnamefont {T.}~\bibnamefont
  {Fujita}}\ and\ \bibinfo {author} {\bibfnamefont {R.}~\bibnamefont
  {Durrer}},\ }\href@noop {} {\  (\bibinfo {year} {2019})},\ \Eprint
  {http://arxiv.org/abs/1904.11428} {arXiv:1904.11428 [astro-ph.CO]}
  \BibitemShut {NoStop}%
\bibitem [{\citenamefont {{Vachaspati}}(1991)}]{vachaspati}%
  \BibitemOpen
  \bibfield  {author} {\bibinfo {author} {\bibfnamefont {T.}~\bibnamefont
  {{Vachaspati}}},\ }\href {\doibase 10.1016/0370-2693(91)90051-Q} {\bibfield
  {journal} {\bibinfo  {journal} {Physics Letters B}\ }\textbf {\bibinfo
  {volume} {265}},\ \bibinfo {pages} {258} (\bibinfo {year}
  {1991})}\BibitemShut {NoStop}%
\bibitem [{\citenamefont {Sigl}\ \emph {et~al.}(1997)\citenamefont {Sigl},
  \citenamefont {Olinto},\ and\ \citenamefont {Jedamzik}}]{Sigl:1996dm}%
  \BibitemOpen
  \bibfield  {author} {\bibinfo {author} {\bibfnamefont {G.}~\bibnamefont
  {Sigl}}, \bibinfo {author} {\bibfnamefont {A.~V.}\ \bibnamefont {Olinto}}, \
  and\ \bibinfo {author} {\bibfnamefont {K.}~\bibnamefont {Jedamzik}},\ }\href
  {\doibase 10.1103/PhysRevD.55.4582} {\bibfield  {journal} {\bibinfo
  {journal} {Phys. Rev.}\ }\textbf {\bibinfo {volume} {D55}},\ \bibinfo {pages}
  {4582} (\bibinfo {year} {1997})},\ \Eprint
  {http://arxiv.org/abs/astro-ph/9610201} {arXiv:astro-ph/9610201 [astro-ph]}
  \BibitemShut {NoStop}%
\bibitem [{\citenamefont {{Kisslinger}}(2003)}]{kisslinger}%
  \BibitemOpen
  \bibfield  {author} {\bibinfo {author} {\bibfnamefont {L.~S.}\ \bibnamefont
  {{Kisslinger}}},\ }\href {\doibase 10.1103/PhysRevD.68.043516} {\bibfield
  {journal} {\bibinfo  {journal} {\prd}\ }\textbf {\bibinfo {volume} {68}},\
  \bibinfo {eid} {043516} (\bibinfo {year} {2003})},\ \Eprint
  {http://arxiv.org/abs/hep-ph/0212206} {hep-ph/0212206} \BibitemShut {NoStop}%
\bibitem [{\citenamefont {{Tevzadze}}\ \emph {et~al.}(2012)\citenamefont
  {{Tevzadze}}, \citenamefont {{Kisslinger}}, \citenamefont {{Brandenburg}},\
  and\ \citenamefont {{Kahniashvili}}}]{qcd}%
  \BibitemOpen
  \bibfield  {author} {\bibinfo {author} {\bibfnamefont {A.~G.}\ \bibnamefont
  {{Tevzadze}}}, \bibinfo {author} {\bibfnamefont {L.}~\bibnamefont
  {{Kisslinger}}}, \bibinfo {author} {\bibfnamefont {A.}~\bibnamefont
  {{Brandenburg}}}, \ and\ \bibinfo {author} {\bibfnamefont {T.}~\bibnamefont
  {{Kahniashvili}}},\ }\href {\doibase 10.1088/0004-637X/759/1/54} {\bibfield
  {journal} {\bibinfo  {journal} {Astrophys. J.}\ }\textbf {\bibinfo {volume}
  {759}},\ \bibinfo {eid} {54} (\bibinfo {year} {2012})},\ \Eprint
  {http://arxiv.org/abs/1207.0751} {arXiv:1207.0751 [astro-ph.CO]} \BibitemShut
  {NoStop}%
\bibitem [{\citenamefont {{Biermann}}(1950)}]{biermann}%
  \BibitemOpen
  \bibfield  {author} {\bibinfo {author} {\bibfnamefont {L.}~\bibnamefont
  {{Biermann}}},\ }\href@noop {} {\bibfield  {journal} {\bibinfo  {journal}
  {Zeitschrift Naturforschung Teil A}\ }\textbf {\bibinfo {volume} {5}},\
  \bibinfo {pages} {65} (\bibinfo {year} {1950})}\BibitemShut {NoStop}%
\bibitem [{\citenamefont {Fenu}\ \emph {et~al.}(2011)\citenamefont {Fenu},
  \citenamefont {Pitrou},\ and\ \citenamefont {Maartens}}]{fenu}%
  \BibitemOpen
  \bibfield  {author} {\bibinfo {author} {\bibfnamefont {E.}~\bibnamefont
  {Fenu}}, \bibinfo {author} {\bibfnamefont {C.}~\bibnamefont {Pitrou}}, \ and\
  \bibinfo {author} {\bibfnamefont {R.}~\bibnamefont {Maartens}},\ }\href
  {\doibase 10.1111/j.1365-2966.2011.18554.x} {\bibfield  {journal} {\bibinfo
  {journal} {Mon. Not. Roy. Astron. Soc.}\ }\textbf {\bibinfo {volume} {414}},\
  \bibinfo {pages} {2354} (\bibinfo {year} {2011})},\ \Eprint
  {http://arxiv.org/abs/1012.2958} {arXiv:1012.2958 [astro-ph.CO]} \BibitemShut
  {NoStop}%
\bibitem [{\citenamefont {{Subramanian}}\ \emph {et~al.}(1994)\citenamefont
  {{Subramanian}}, \citenamefont {{Narasimha}},\ and\ \citenamefont
  {{Chitre}}}]{kandu1994}%
  \BibitemOpen
  \bibfield  {author} {\bibinfo {author} {\bibfnamefont {K.}~\bibnamefont
  {{Subramanian}}}, \bibinfo {author} {\bibfnamefont {D.}~\bibnamefont
  {{Narasimha}}}, \ and\ \bibinfo {author} {\bibfnamefont {S.~M.}\ \bibnamefont
  {{Chitre}}},\ }\href {\doibase 10.1093/mnras/271.1.L15} {\bibfield  {journal}
  {\bibinfo  {journal} {Mon. Not. Roy. Astron. Soc.}\ }\textbf {\bibinfo
  {volume} {271}} (\bibinfo {year} {1994}),\
  10.1093/mnras/271.1.L15}\BibitemShut {NoStop}%
\bibitem [{\citenamefont {{Gnedin}}\ \emph {et~al.}(2000)\citenamefont
  {{Gnedin}}, \citenamefont {{Ferrara}},\ and\ \citenamefont
  {{Zweibel}}}]{Zweibel2000}%
  \BibitemOpen
  \bibfield  {author} {\bibinfo {author} {\bibfnamefont {N.~Y.}\ \bibnamefont
  {{Gnedin}}}, \bibinfo {author} {\bibfnamefont {A.}~\bibnamefont {{Ferrara}}},
  \ and\ \bibinfo {author} {\bibfnamefont {E.~G.}\ \bibnamefont {{Zweibel}}},\
  }\href {\doibase 10.1086/309272} {\bibfield  {journal} {\bibinfo  {journal}
  {\apj}\ }\textbf {\bibinfo {volume} {539}},\ \bibinfo {pages} {505} (\bibinfo
  {year} {2000})},\ \Eprint {http://arxiv.org/abs/astro-ph/0001066}
  {astro-ph/0001066} \BibitemShut {NoStop}%
\bibitem [{\citenamefont {Kulsrud}\ \emph {et~al.}(1997)\citenamefont
  {Kulsrud}, \citenamefont {Cen}, \citenamefont {Ostriker},\ and\ \citenamefont
  {Ryu}}]{kulsrud1997}%
  \BibitemOpen
  \bibfield  {author} {\bibinfo {author} {\bibfnamefont {R.~M.}\ \bibnamefont
  {Kulsrud}}, \bibinfo {author} {\bibfnamefont {R.}~\bibnamefont {Cen}},
  \bibinfo {author} {\bibfnamefont {J.~P.}\ \bibnamefont {Ostriker}}, \ and\
  \bibinfo {author} {\bibfnamefont {D.}~\bibnamefont {Ryu}},\ }\href {\doibase
  10.1086/303987} {\bibfield  {journal} {\bibinfo  {journal} {Astrophys. J.}\
  }\textbf {\bibinfo {volume} {480}},\ \bibinfo {pages} {481} (\bibinfo {year}
  {1997})},\ \Eprint {http://arxiv.org/abs/astro-ph/9607141}
  {arXiv:astro-ph/9607141 [astro-ph]} \BibitemShut {NoStop}%
\bibitem [{\citenamefont {{Demozzi}}\ \emph {et~al.}(2009)\citenamefont
  {{Demozzi}}, \citenamefont {{Mukhanov}},\ and\ \citenamefont
  {{Rubinstein}}}]{mukhanov2009}%
  \BibitemOpen
  \bibfield  {author} {\bibinfo {author} {\bibfnamefont {V.}~\bibnamefont
  {{Demozzi}}}, \bibinfo {author} {\bibfnamefont {V.}~\bibnamefont
  {{Mukhanov}}}, \ and\ \bibinfo {author} {\bibfnamefont {H.}~\bibnamefont
  {{Rubinstein}}},\ }\href {\doibase 10.1088/1475-7516/2009/08/025} {\bibfield
  {journal} {\bibinfo  {journal} {Journal of Cosmology and Astroparticle
  Physics}\ }\textbf {\bibinfo {volume} {8}},\ \bibinfo {eid} {025} (\bibinfo
  {year} {2009})},\ \Eprint {http://arxiv.org/abs/0907.1030} {arXiv:0907.1030
  [astro-ph.CO]} \BibitemShut {NoStop}%
\bibitem [{\citenamefont {{Kobayashi}}\ and\ \citenamefont
  {{Afshordi}}(2014)}]{kobayashi:2014}%
  \BibitemOpen
  \bibfield  {author} {\bibinfo {author} {\bibfnamefont {T.}~\bibnamefont
  {{Kobayashi}}}\ and\ \bibinfo {author} {\bibfnamefont {N.}~\bibnamefont
  {{Afshordi}}},\ }\href {\doibase 10.1007/JHEP10(2014)166} {\bibfield
  {journal} {\bibinfo  {journal} {Journal of High Energy Physics}\ }\textbf
  {\bibinfo {volume} {10}},\ \bibinfo {eid} {166} (\bibinfo {year} {2014})},\
  \Eprint {http://arxiv.org/abs/1408.4141} {arXiv:1408.4141 [hep-th]}
  \BibitemShut {NoStop}%
\bibitem [{\citenamefont {Sharma}\ \emph {et~al.}(2018)\citenamefont {Sharma},
  \citenamefont {Subramanian},\ and\ \citenamefont {Seshadri}}]{sharmahelical}%
  \BibitemOpen
  \bibfield  {author} {\bibinfo {author} {\bibfnamefont {R.}~\bibnamefont
  {Sharma}}, \bibinfo {author} {\bibfnamefont {K.}~\bibnamefont {Subramanian}},
  \ and\ \bibinfo {author} {\bibfnamefont {T.~R.}\ \bibnamefont {Seshadri}},\
  }\href {\doibase 10.1103/PhysRevD.97.083503} {\bibfield  {journal} {\bibinfo
  {journal} {Phys. Rev.}\ }\textbf {\bibinfo {volume} {D97}},\ \bibinfo {pages}
  {083503} (\bibinfo {year} {2018})},\ \Eprint
  {http://arxiv.org/abs/1802.04847} {arXiv:1802.04847 [astro-ph.CO]}
  \BibitemShut {NoStop}%
\bibitem [{\citenamefont {Arzoumanian}\ \emph
  {et~al.}(2020{\natexlab{a}})\citenamefont {Arzoumanian} \emph
  {et~al.}}]{nanograv}%
  \BibitemOpen
  \bibfield  {author} {\bibinfo {author} {\bibfnamefont {Z.}~\bibnamefont
  {Arzoumanian}} \emph {et~al.} (\bibinfo {collaboration} {NANOGrav}),\
  }\href@noop {} {\  (\bibinfo {year} {2020}{\natexlab{a}})},\ \Eprint
  {http://arxiv.org/abs/2009.04496} {arXiv:2009.04496 [astro-ph.HE]}
  \BibitemShut {NoStop}%
\bibitem [{\citenamefont {Ding}\ \emph {et~al.}(2020)\citenamefont {Ding},
  \citenamefont {Tong},\ and\ \citenamefont {Wang}}]{nanogavastro1}%
  \BibitemOpen
  \bibfield  {author} {\bibinfo {author} {\bibfnamefont {Q.}~\bibnamefont
  {Ding}}, \bibinfo {author} {\bibfnamefont {X.}~\bibnamefont {Tong}}, \ and\
  \bibinfo {author} {\bibfnamefont {Y.}~\bibnamefont {Wang}},\ }\href@noop {}
  {\  (\bibinfo {year} {2020})},\ \Eprint {http://arxiv.org/abs/2009.11106}
  {arXiv:2009.11106 [astro-ph.HE]} \BibitemShut {NoStop}%
\bibitem [{\citenamefont {Arzoumanian}\ \emph
  {et~al.}(2020{\natexlab{b}})\citenamefont {Arzoumanian} \emph
  {et~al.}}]{nanogravastro2}%
  \BibitemOpen
  \bibfield  {author} {\bibinfo {author} {\bibfnamefont {Z.}~\bibnamefont
  {Arzoumanian}} \emph {et~al.} (\bibinfo {collaboration} {NANOGrav}),\
  }\href@noop {} {\  (\bibinfo {year} {2020}{\natexlab{b}})},\ \Eprint
  {http://arxiv.org/abs/2009.04496} {arXiv:2009.04496 [astro-ph.HE]}
  \BibitemShut {NoStop}%
\bibitem [{\citenamefont {Ellis}\ and\ \citenamefont
  {Lewicki}(2020)}]{nanogravstring1}%
  \BibitemOpen
  \bibfield  {author} {\bibinfo {author} {\bibfnamefont {J.}~\bibnamefont
  {Ellis}}\ and\ \bibinfo {author} {\bibfnamefont {M.}~\bibnamefont
  {Lewicki}},\ }\href@noop {} {\  (\bibinfo {year} {2020})},\ \Eprint
  {http://arxiv.org/abs/2009.06555} {arXiv:2009.06555 [astro-ph.CO]}
  \BibitemShut {NoStop}%
\bibitem [{\citenamefont {Blasi}\ \emph {et~al.}(2020)\citenamefont {Blasi},
  \citenamefont {Brdar},\ and\ \citenamefont {Schmitz}}]{nanogravstring2}%
  \BibitemOpen
  \bibfield  {author} {\bibinfo {author} {\bibfnamefont {S.}~\bibnamefont
  {Blasi}}, \bibinfo {author} {\bibfnamefont {V.}~\bibnamefont {Brdar}}, \ and\
  \bibinfo {author} {\bibfnamefont {K.}~\bibnamefont {Schmitz}},\ }\href@noop
  {} {\  (\bibinfo {year} {2020})},\ \Eprint {http://arxiv.org/abs/2009.06607}
  {arXiv:2009.06607 [astro-ph.CO]} \BibitemShut {NoStop}%
\bibitem [{\citenamefont {Buchmuller}\ \emph {et~al.}(2020)\citenamefont
  {Buchmuller}, \citenamefont {Domcke},\ and\ \citenamefont
  {Schmitz}}]{nanogravstring3}%
  \BibitemOpen
  \bibfield  {author} {\bibinfo {author} {\bibfnamefont {W.}~\bibnamefont
  {Buchmuller}}, \bibinfo {author} {\bibfnamefont {V.}~\bibnamefont {Domcke}},
  \ and\ \bibinfo {author} {\bibfnamefont {K.}~\bibnamefont {Schmitz}},\
  }\href@noop {} {\  (\bibinfo {year} {2020})},\ \Eprint
  {http://arxiv.org/abs/2009.10649} {arXiv:2009.10649 [astro-ph.CO]}
  \BibitemShut {NoStop}%
\bibitem [{\citenamefont {Bian}\ \emph {et~al.}(2020)\citenamefont {Bian},
  \citenamefont {Cai}, \citenamefont {Liu}, \citenamefont {Yang},\ and\
  \citenamefont {Zhou}}]{bian2020}%
  \BibitemOpen
  \bibfield  {author} {\bibinfo {author} {\bibfnamefont {L.}~\bibnamefont
  {Bian}}, \bibinfo {author} {\bibfnamefont {R.-G.}\ \bibnamefont {Cai}},
  \bibinfo {author} {\bibfnamefont {J.}~\bibnamefont {Liu}}, \bibinfo {author}
  {\bibfnamefont {X.-Y.}\ \bibnamefont {Yang}}, \ and\ \bibinfo {author}
  {\bibfnamefont {R.}~\bibnamefont {Zhou}},\ }\href@noop {} {\  (\bibinfo
  {year} {2020})},\ \Eprint {http://arxiv.org/abs/2009.13893} {arXiv:2009.13893
  [astro-ph.CO]} \BibitemShut {NoStop}%
\bibitem [{\citenamefont {Vaskonen}\ and\ \citenamefont
  {Veerm\"ae}(2020)}]{nanopbh1}%
  \BibitemOpen
  \bibfield  {author} {\bibinfo {author} {\bibfnamefont {V.}~\bibnamefont
  {Vaskonen}}\ and\ \bibinfo {author} {\bibfnamefont {H.}~\bibnamefont
  {Veerm\"ae}},\ }\href@noop {} {\  (\bibinfo {year} {2020})},\ \Eprint
  {http://arxiv.org/abs/2009.07832} {arXiv:2009.07832 [astro-ph.CO]}
  \BibitemShut {NoStop}%
\bibitem [{\citenamefont {Kohri}\ and\ \citenamefont
  {Terada}(2020)}]{nanopbh2}%
  \BibitemOpen
  \bibfield  {author} {\bibinfo {author} {\bibfnamefont {K.}~\bibnamefont
  {Kohri}}\ and\ \bibinfo {author} {\bibfnamefont {T.}~\bibnamefont {Terada}},\
  }\href@noop {} {\  (\bibinfo {year} {2020})},\ \Eprint
  {http://arxiv.org/abs/2009.11853} {arXiv:2009.11853 [astro-ph.CO]}
  \BibitemShut {NoStop}%
\bibitem [{\citenamefont {Bhaumik}\ and\ \citenamefont
  {Jain}(2020)}]{nanopbh3}%
  \BibitemOpen
  \bibfield  {author} {\bibinfo {author} {\bibfnamefont {N.}~\bibnamefont
  {Bhaumik}}\ and\ \bibinfo {author} {\bibfnamefont {R.~K.}\ \bibnamefont
  {Jain}},\ }\href@noop {} {\  (\bibinfo {year} {2020})},\ \Eprint
  {http://arxiv.org/abs/2009.10424} {arXiv:2009.10424 [astro-ph.CO]}
  \BibitemShut {NoStop}%
\bibitem [{\citenamefont {De~Luca}\ \emph {et~al.}(2020)\citenamefont
  {De~Luca}, \citenamefont {Franciolini},\ and\ \citenamefont
  {Riotto}}]{nanopbh4}%
  \BibitemOpen
  \bibfield  {author} {\bibinfo {author} {\bibfnamefont {V.}~\bibnamefont
  {De~Luca}}, \bibinfo {author} {\bibfnamefont {G.}~\bibnamefont
  {Franciolini}}, \ and\ \bibinfo {author} {\bibfnamefont {A.}~\bibnamefont
  {Riotto}},\ }\href@noop {} {\  (\bibinfo {year} {2020})},\ \Eprint
  {http://arxiv.org/abs/2009.08268} {arXiv:2009.08268 [astro-ph.CO]}
  \BibitemShut {NoStop}%
\bibitem [{\citenamefont {Kitajima}\ \emph {et~al.}(2020)\citenamefont
  {Kitajima}, \citenamefont {Soda},\ and\ \citenamefont {Urakawa}}]{soda2020}%
  \BibitemOpen
  \bibfield  {author} {\bibinfo {author} {\bibfnamefont {N.}~\bibnamefont
  {Kitajima}}, \bibinfo {author} {\bibfnamefont {J.}~\bibnamefont {Soda}}, \
  and\ \bibinfo {author} {\bibfnamefont {Y.}~\bibnamefont {Urakawa}},\
  }\href@noop {} {\  (\bibinfo {year} {2020})},\ \Eprint
  {http://arxiv.org/abs/2010.10990} {arXiv:2010.10990 [astro-ph.CO]}
  \BibitemShut {NoStop}%
\bibitem [{\citenamefont {Dom\`enech}\ and\ \citenamefont
  {Pi}(2020)}]{domenech2020}%
  \BibitemOpen
  \bibfield  {author} {\bibinfo {author} {\bibfnamefont {G.}~\bibnamefont
  {Dom\`enech}}\ and\ \bibinfo {author} {\bibfnamefont {S.}~\bibnamefont
  {Pi}},\ }\href@noop {} {\  (\bibinfo {year} {2020})},\ \Eprint
  {http://arxiv.org/abs/2010.03976} {arXiv:2010.03976 [astro-ph.CO]}
  \BibitemShut {NoStop}%
\bibitem [{\citenamefont {Lewicki}\ and\ \citenamefont
  {Vaskonen}(2020)}]{marek2020dec}%
  \BibitemOpen
  \bibfield  {author} {\bibinfo {author} {\bibfnamefont {M.}~\bibnamefont
  {Lewicki}}\ and\ \bibinfo {author} {\bibfnamefont {V.}~\bibnamefont
  {Vaskonen}},\ }\href@noop {} {\  (\bibinfo {year} {2020})},\ \Eprint
  {http://arxiv.org/abs/2012.07826} {arXiv:2012.07826 [astro-ph.CO]}
  \BibitemShut {NoStop}%
\bibitem [{\citenamefont {Addazi}\ \emph {et~al.}(2020)\citenamefont {Addazi},
  \citenamefont {Cai}, \citenamefont {Gan}, \citenamefont {Marciano},\ and\
  \citenamefont {Zeng}}]{addazi2020}%
  \BibitemOpen
  \bibfield  {author} {\bibinfo {author} {\bibfnamefont {A.}~\bibnamefont
  {Addazi}}, \bibinfo {author} {\bibfnamefont {Y.-F.}\ \bibnamefont {Cai}},
  \bibinfo {author} {\bibfnamefont {Q.}~\bibnamefont {Gan}}, \bibinfo {author}
  {\bibfnamefont {A.}~\bibnamefont {Marciano}}, \ and\ \bibinfo {author}
  {\bibfnamefont {K.}~\bibnamefont {Zeng}},\ }\href@noop {} {\  (\bibinfo
  {year} {2020})},\ \Eprint {http://arxiv.org/abs/2009.10327} {arXiv:2009.10327
  [hep-ph]} \BibitemShut {NoStop}%
\bibitem [{\citenamefont {Nakai}\ \emph {et~al.}(2020)\citenamefont {Nakai},
  \citenamefont {Suzuki}, \citenamefont {Takahashi},\ and\ \citenamefont
  {Yamada}}]{nakai2020}%
  \BibitemOpen
  \bibfield  {author} {\bibinfo {author} {\bibfnamefont {Y.}~\bibnamefont
  {Nakai}}, \bibinfo {author} {\bibfnamefont {M.}~\bibnamefont {Suzuki}},
  \bibinfo {author} {\bibfnamefont {F.}~\bibnamefont {Takahashi}}, \ and\
  \bibinfo {author} {\bibfnamefont {M.}~\bibnamefont {Yamada}},\ }\href@noop {}
  {\  (\bibinfo {year} {2020})},\ \Eprint {http://arxiv.org/abs/2009.09754}
  {arXiv:2009.09754 [astro-ph.CO]} \BibitemShut {NoStop}%
\bibitem [{\citenamefont {Neronov}\ \emph {et~al.}(2020)\citenamefont
  {Neronov}, \citenamefont {Roper~Pol}, \citenamefont {Caprini},\ and\
  \citenamefont {Semikoz}}]{caprini2020}%
  \BibitemOpen
  \bibfield  {author} {\bibinfo {author} {\bibfnamefont {A.}~\bibnamefont
  {Neronov}}, \bibinfo {author} {\bibfnamefont {A.}~\bibnamefont {Roper~Pol}},
  \bibinfo {author} {\bibfnamefont {C.}~\bibnamefont {Caprini}}, \ and\
  \bibinfo {author} {\bibfnamefont {D.}~\bibnamefont {Semikoz}},\ }\href@noop
  {} {\  (\bibinfo {year} {2020})},\ \Eprint {http://arxiv.org/abs/2009.14174}
  {arXiv:2009.14174 [astro-ph.CO]} \BibitemShut {NoStop}%
\bibitem [{\citenamefont {Samanta}\ and\ \citenamefont
  {Datta}(2020)}]{samanta2020}%
  \BibitemOpen
  \bibfield  {author} {\bibinfo {author} {\bibfnamefont {R.}~\bibnamefont
  {Samanta}}\ and\ \bibinfo {author} {\bibfnamefont {S.}~\bibnamefont
  {Datta}},\ }\href@noop {} {\  (\bibinfo {year} {2020})},\ \Eprint
  {http://arxiv.org/abs/2009.13452} {arXiv:2009.13452 [hep-ph]} \BibitemShut
  {NoStop}%
\bibitem [{\citenamefont {Ratzinger}\ and\ \citenamefont
  {Schwaller}(2020)}]{ratzinger2020}%
  \BibitemOpen
  \bibfield  {author} {\bibinfo {author} {\bibfnamefont {W.}~\bibnamefont
  {Ratzinger}}\ and\ \bibinfo {author} {\bibfnamefont {P.}~\bibnamefont
  {Schwaller}},\ }\href@noop {} {\  (\bibinfo {year} {2020})},\ \Eprint
  {http://arxiv.org/abs/2009.11875} {arXiv:2009.11875 [astro-ph.CO]}
  \BibitemShut {NoStop}%
\bibitem [{\citenamefont {Vagnozzi}(2021)}]{vagnozzi2020}%
  \BibitemOpen
  \bibfield  {author} {\bibinfo {author} {\bibfnamefont {S.}~\bibnamefont
  {Vagnozzi}},\ }\href {\doibase 10.1093/mnrasl/slaa203} {\bibfield  {journal}
  {\bibinfo  {journal} {Mon. Not. Roy. Astron. Soc.}\ }\textbf {\bibinfo
  {volume} {502}},\ \bibinfo {pages} {L11} (\bibinfo {year} {2021})},\ \Eprint
  {http://arxiv.org/abs/2009.13432} {arXiv:2009.13432 [astro-ph.CO]}
  \BibitemShut {NoStop}%
\bibitem [{\citenamefont {Pandey}(2020)}]{pandey2020}%
  \BibitemOpen
  \bibfield  {author} {\bibinfo {author} {\bibfnamefont {A.~K.}\ \bibnamefont
  {Pandey}},\ }\href@noop {} {\  (\bibinfo {year} {2020})},\ \Eprint
  {http://arxiv.org/abs/2011.05821} {arXiv:2011.05821 [astro-ph.CO]}
  \BibitemShut {NoStop}%
\bibitem [{\citenamefont {Ramberg}\ and\ \citenamefont
  {Visinelli}(2020)}]{ramberg2020}%
  \BibitemOpen
  \bibfield  {author} {\bibinfo {author} {\bibfnamefont {N.}~\bibnamefont
  {Ramberg}}\ and\ \bibinfo {author} {\bibfnamefont {L.}~\bibnamefont
  {Visinelli}}\ }(\bibinfo {year} {2020})\ \Eprint
  {http://arxiv.org/abs/2012.06882} {arXiv:2012.06882 [astro-ph.CO]}
  \BibitemShut {NoStop}%
\bibitem [{\citenamefont {Bhattacharya}\ \emph {et~al.}(2020)\citenamefont
  {Bhattacharya}, \citenamefont {Mohanty},\ and\ \citenamefont
  {Parashari}}]{bhattacharya2020}%
  \BibitemOpen
  \bibfield  {author} {\bibinfo {author} {\bibfnamefont {S.}~\bibnamefont
  {Bhattacharya}}, \bibinfo {author} {\bibfnamefont {S.}~\bibnamefont
  {Mohanty}}, \ and\ \bibinfo {author} {\bibfnamefont {P.}~\bibnamefont
  {Parashari}},\ }\href@noop {} {\  (\bibinfo {year} {2020})},\ \Eprint
  {http://arxiv.org/abs/2010.05071} {arXiv:2010.05071 [astro-ph.CO]}
  \BibitemShut {NoStop}%
\bibitem [{\citenamefont {Sharma}\ \emph {et~al.}(2020)\citenamefont {Sharma},
  \citenamefont {Subramanian},\ and\ \citenamefont {Seshadri}}]{sharmagw}%
  \BibitemOpen
  \bibfield  {author} {\bibinfo {author} {\bibfnamefont {R.}~\bibnamefont
  {Sharma}}, \bibinfo {author} {\bibfnamefont {K.}~\bibnamefont {Subramanian}},
  \ and\ \bibinfo {author} {\bibfnamefont {T.~R.}\ \bibnamefont {Seshadri}},\
  }\href {\doibase 10.1103/PhysRevD.101.103526} {\bibfield  {journal} {\bibinfo
   {journal} {Phys. Rev. D}\ }\textbf {\bibinfo {volume} {101}},\ \bibinfo
  {pages} {103526} (\bibinfo {year} {2020})}\BibitemShut {NoStop}%
\bibitem [{\citenamefont {Moore}\ \emph {et~al.}(2015)\citenamefont {Moore},
  \citenamefont {Cole},\ and\ \citenamefont {Berry}}]{moore2014}%
  \BibitemOpen
  \bibfield  {author} {\bibinfo {author} {\bibfnamefont {C.}~\bibnamefont
  {Moore}}, \bibinfo {author} {\bibfnamefont {R.}~\bibnamefont {Cole}}, \ and\
  \bibinfo {author} {\bibfnamefont {C.}~\bibnamefont {Berry}},\ }\href
  {\doibase 10.1088/0264-9381/32/1/015014} {\bibfield  {journal} {\bibinfo
  {journal} {Class. Quant. Grav.}\ }\textbf {\bibinfo {volume} {32}},\ \bibinfo
  {pages} {015014} (\bibinfo {year} {2015})},\ \Eprint
  {http://arxiv.org/abs/1408.0740} {arXiv:1408.0740 [gr-qc]} \BibitemShut
  {NoStop}%
\bibitem [{\citenamefont {{Brandenburg}}\ \emph {et~al.}(2015)\citenamefont
  {{Brandenburg}}, \citenamefont {{Kahniashvili}},\ and\ \citenamefont
  {{Tevzadze}}}]{axel}%
  \BibitemOpen
  \bibfield  {author} {\bibinfo {author} {\bibfnamefont {A.}~\bibnamefont
  {{Brandenburg}}}, \bibinfo {author} {\bibfnamefont {T.}~\bibnamefont
  {{Kahniashvili}}}, \ and\ \bibinfo {author} {\bibfnamefont {A.~G.}\
  \bibnamefont {{Tevzadze}}},\ }\href {\doibase 10.1103/PhysRevLett.114.075001}
  {\bibfield  {journal} {\bibinfo  {journal} {Physical Review Letters}\
  }\textbf {\bibinfo {volume} {114}},\ \bibinfo {eid} {075001} (\bibinfo {year}
  {2015})},\ \Eprint {http://arxiv.org/abs/1404.2238} {arXiv:1404.2238}
  \BibitemShut {NoStop}%
\bibitem [{\citenamefont {Brandenburg}\ and\ \citenamefont
  {Kahniashvili}(2017)}]{axel2017}%
  \BibitemOpen
  \bibfield  {author} {\bibinfo {author} {\bibfnamefont {A.}~\bibnamefont
  {Brandenburg}}\ and\ \bibinfo {author} {\bibfnamefont {T.}~\bibnamefont
  {Kahniashvili}},\ }\href {\doibase 10.1103/PhysRevLett.118.055102} {\bibfield
   {journal} {\bibinfo  {journal} {Phys. Rev. Lett.}\ }\textbf {\bibinfo
  {volume} {118}},\ \bibinfo {pages} {055102} (\bibinfo {year} {2017})},\
  \Eprint {http://arxiv.org/abs/1607.01360} {arXiv:1607.01360
  [physics.flu-dyn]} \BibitemShut {NoStop}%
\bibitem [{\citenamefont {Zrake}(2014)}]{zrake}%
  \BibitemOpen
  \bibfield  {author} {\bibinfo {author} {\bibfnamefont {J.}~\bibnamefont
  {Zrake}},\ }\href {\doibase 10.1088/2041-8205/794/2/L26} {\bibfield
  {journal} {\bibinfo  {journal} {Astrophys. J.}\ }\textbf {\bibinfo {volume}
  {794}},\ \bibinfo {pages} {L26} (\bibinfo {year} {2014})},\ \Eprint
  {http://arxiv.org/abs/1407.5626} {arXiv:1407.5626 [astro-ph.HE]} \BibitemShut
  {NoStop}%
\bibitem [{\citenamefont {{Banerjee}}\ and\ \citenamefont
  {{Jedamzik}}(2004)}]{jedamzik}%
  \BibitemOpen
  \bibfield  {author} {\bibinfo {author} {\bibfnamefont {R.}~\bibnamefont
  {{Banerjee}}}\ and\ \bibinfo {author} {\bibfnamefont {K.}~\bibnamefont
  {{Jedamzik}}},\ }\href {\doibase 10.1103/PhysRevD.70.123003} {\bibfield
  {journal} {\bibinfo  {journal} {\prd}\ }\textbf {\bibinfo {volume} {70}},\
  \bibinfo {eid} {123003} (\bibinfo {year} {2004})},\ \Eprint
  {http://arxiv.org/abs/astro-ph/0410032} {astro-ph/0410032} \BibitemShut
  {NoStop}%
\bibitem [{\citenamefont {{Kahniashvili}}\ \emph {et~al.}(2013)\citenamefont
  {{Kahniashvili}}, \citenamefont {{Tevzadze}}, \citenamefont {{Brandenburg}},\
  and\ \citenamefont {{Neronov}}}]{tina2013}%
  \BibitemOpen
  \bibfield  {author} {\bibinfo {author} {\bibfnamefont {T.}~\bibnamefont
  {{Kahniashvili}}}, \bibinfo {author} {\bibfnamefont {A.~G.}\ \bibnamefont
  {{Tevzadze}}}, \bibinfo {author} {\bibfnamefont {A.}~\bibnamefont
  {{Brandenburg}}}, \ and\ \bibinfo {author} {\bibfnamefont {A.}~\bibnamefont
  {{Neronov}}},\ }\href {\doibase 10.1103/PhysRevD.87.083007} {\bibfield
  {journal} {\bibinfo  {journal} {\prd}\ }\textbf {\bibinfo {volume} {87}},\
  \bibinfo {eid} {083007} (\bibinfo {year} {2013})},\ \Eprint
  {http://arxiv.org/abs/1212.0596} {arXiv:1212.0596 [astro-ph.CO]} \BibitemShut
  {NoStop}%
\bibitem [{\citenamefont {Kahniashvili}\ \emph {et~al.}(2016)\citenamefont
  {Kahniashvili}, \citenamefont {Brandenburg},\ and\ \citenamefont
  {Tevzadze}}]{axel2016}%
  \BibitemOpen
  \bibfield  {author} {\bibinfo {author} {\bibfnamefont {T.}~\bibnamefont
  {Kahniashvili}}, \bibinfo {author} {\bibfnamefont {A.}~\bibnamefont
  {Brandenburg}}, \ and\ \bibinfo {author} {\bibfnamefont {A.~G.}\ \bibnamefont
  {Tevzadze}},\ }\bibfield  {booktitle} {\emph {\bibinfo {booktitle} {{ICTP
  Summer School on Cosmology 2014 Trieste, Italy, August 4-21, 2014}}},\ }\href
  {\doibase 10.1088/0031-8949/91/10/104008} {\bibfield  {journal} {\bibinfo
  {journal} {Phys. Scripta}\ }\textbf {\bibinfo {volume} {91}},\ \bibinfo
  {pages} {104008} (\bibinfo {year} {2016})},\ \Eprint
  {http://arxiv.org/abs/1507.00510} {arXiv:1507.00510 [astro-ph.CO]}
  \BibitemShut {NoStop}%
\bibitem [{\citenamefont {Trivedi}\ \emph {et~al.}(2014)\citenamefont
  {Trivedi}, \citenamefont {Subramanian},\ and\ \citenamefont
  {Seshadri}}]{trivedi2013}%
  \BibitemOpen
  \bibfield  {author} {\bibinfo {author} {\bibfnamefont {P.}~\bibnamefont
  {Trivedi}}, \bibinfo {author} {\bibfnamefont {K.}~\bibnamefont
  {Subramanian}}, \ and\ \bibinfo {author} {\bibfnamefont {T.}~\bibnamefont
  {Seshadri}},\ }\href {\doibase 10.1103/PhysRevD.89.043523} {\bibfield
  {journal} {\bibinfo  {journal} {Phys. Rev. D}\ }\textbf {\bibinfo {volume}
  {89}},\ \bibinfo {pages} {043523} (\bibinfo {year} {2014})},\ \Eprint
  {http://arxiv.org/abs/1312.5308} {arXiv:1312.5308 [astro-ph.CO]} \BibitemShut
  {NoStop}%
\bibitem [{\citenamefont {{de Salas}}\ \emph {et~al.}(2015)\citenamefont {{de
  Salas}}, \citenamefont {{Lattanzi}}, \citenamefont {{Mangano}}, \citenamefont
  {{Miele}}, \citenamefont {{Pastor}},\ and\ \citenamefont {{Pisanti}}}]{bbn}%
  \BibitemOpen
  \bibfield  {author} {\bibinfo {author} {\bibfnamefont {P.~F.}\ \bibnamefont
  {{de Salas}}}, \bibinfo {author} {\bibfnamefont {M.}~\bibnamefont
  {{Lattanzi}}}, \bibinfo {author} {\bibfnamefont {G.}~\bibnamefont
  {{Mangano}}}, \bibinfo {author} {\bibfnamefont {G.}~\bibnamefont {{Miele}}},
  \bibinfo {author} {\bibfnamefont {S.}~\bibnamefont {{Pastor}}}, \ and\
  \bibinfo {author} {\bibfnamefont {O.}~\bibnamefont {{Pisanti}}},\ }\href
  {\doibase 10.1103/PhysRevD.92.123534} {\bibfield  {journal} {\bibinfo
  {journal} {\prd}\ }\textbf {\bibinfo {volume} {92}},\ \bibinfo {eid} {123534}
  (\bibinfo {year} {2015})},\ \Eprint {http://arxiv.org/abs/1511.00672}
  {arXiv:1511.00672} \BibitemShut {NoStop}%
\bibitem [{\citenamefont {Bartolo}\ \emph {et~al.}(2018)\citenamefont
  {Bartolo}, \citenamefont {Domcke}, \citenamefont {Figueroa}, \citenamefont
  {Garc\'\i{}a-Bellido}, \citenamefont {Peloso}, \citenamefont {Pieroni},
  \citenamefont {Ricciardone}, \citenamefont {Sakellariadou}, \citenamefont
  {Sorbo},\ and\ \citenamefont {Tasinato}}]{bartolo2018}%
  \BibitemOpen
  \bibfield  {author} {\bibinfo {author} {\bibfnamefont {N.}~\bibnamefont
  {Bartolo}}, \bibinfo {author} {\bibfnamefont {V.}~\bibnamefont {Domcke}},
  \bibinfo {author} {\bibfnamefont {D.~G.}\ \bibnamefont {Figueroa}}, \bibinfo
  {author} {\bibfnamefont {J.}~\bibnamefont {Garc\'\i{}a-Bellido}}, \bibinfo
  {author} {\bibfnamefont {M.}~\bibnamefont {Peloso}}, \bibinfo {author}
  {\bibfnamefont {M.}~\bibnamefont {Pieroni}}, \bibinfo {author} {\bibfnamefont
  {A.}~\bibnamefont {Ricciardone}}, \bibinfo {author} {\bibfnamefont
  {M.}~\bibnamefont {Sakellariadou}}, \bibinfo {author} {\bibfnamefont
  {L.}~\bibnamefont {Sorbo}}, \ and\ \bibinfo {author} {\bibfnamefont
  {G.}~\bibnamefont {Tasinato}},\ }\href {\doibase
  10.1088/1475-7516/2018/11/034} {\bibfield  {journal} {\bibinfo  {journal}
  {JCAP}\ }\textbf {\bibinfo {volume} {11}},\ \bibinfo {pages} {034} (\bibinfo
  {year} {2018})},\ \Eprint {http://arxiv.org/abs/1806.02819} {arXiv:1806.02819
  [astro-ph.CO]} \BibitemShut {NoStop}%
\end{thebibliography}%
\end{document}